\def\half{\frac{1}{2}}

\def\bey{\begin{eqnarray}}
\def\eey{\end{eqnarray}}
\def\be{\begin{equation}}
\def\ee{\end{equation}}
\def\ba{\begin{array}}
\def\ea{\end{array}}

\def\sg{\sigma}
\def\Sg{\Sigma}

\def\om{\omega}

\def\pp{\partial}
\def\pp{\partial}

\def\nnb{\nonumber}
\documentclass[twocolumn,showpacs,preprintnumbers,amsmath,amssymb]{revtex4}
\usepackage[colorlinks,linkcolor=blue,citecolor=blue,urlcolor=blue,anchorcolor=blue]{hyperref}
\usepackage{graphicx}% Include figure files
\usepackage{fancyhdr}
\usepackage{dcolumn}% Align table columns on decimal point
\usepackage{bm}
\usepackage{amssymb}
\usepackage{amsmath}
\usepackage{graphicx}
\usepackage[normalem]{ulem}
\usepackage[dvips]{color}

\begin{document}
\preprint{ }

\title{Symmetry energy softening in nuclear matter with non-nucleonic constituents}
\author{Wei-Zhou Jiang\footnote{wzjiang@seu.edu.cn}, Rong-Yao Yang, and Dong-Rui Zhang}
\affiliation{Department of Physics, Southeast University, Nanjing
211189, China }
%\date{}

\begin{abstract}
We study the trend of the nuclear symmetry energy in relativistic
mean-field models with appearance of the hyperon and quark degrees of
freedom at high densities. On the pure hadron level, we focus on the
role of $\Lambda$ hyperons in influencing the symmetry energy both at
given fractions and at charge and chemical equilibriums. The
softening of the nuclear symmetry energy is observed with the
inclusion of the $\Lambda$ hyperons that suppresses the nucleon
fraction. In the phase with the admixture of quarks and hadrons, the
equation of state is established on the Gibbs conditions. With the
increase of the quark volume fraction in denser and denser matter,
the apparent nuclear symmetry energy decreases till to disappear.
This softening would have associations with the observations which
need detailed discriminations in dense matter with the admixture of
new degrees of freedom created by heavy-ion collisions.
\end{abstract}

 \pacs{ 21.65.Ef, 21.60.Jz, 21.65.Qr, 13.75.Ev}
 \keywords{Symmetry energy,
relativistic mean-field models, High-density equation of state }
\maketitle
%\thispagestyle{fancy}
%\pagestyle{fancy}
%\fancyhead[L]{Submitted to Physical Review C}
%\newpage

\section{Introduction}

The nuclear symmetry energy of isospin asymmetric nuclear matter is
not only  important for understanding the structure of neutron- or
proton-rich nuclei and the reaction dynamics of heavy-ion collisions,
see, e.g., Ref.~\cite{ba02,ji05,ba08}, but also plays a crucial role
in a number of important issues in astrophysics, see, e.g.,
Refs.~\cite{lat01,ho01,st10}. Recently, appreciable progresses have
been achieved on constraining the symmetry energy at saturation and
subsaturation densities either through the extraction based on
astrophysical observations or in terms of terrestrial
data~\cite{ch10,Ne2011,St2012a,ts12,Wen2012,Lat2012}. However, the
density dependence of the symmetry energy is still poorly known at
supra-normal densities~\cite{ba08,xiao09,feng10,ru11}. Theoretical
models predict diverse density dependencies of the symmetry energy at
high densities. Noticeably, quite different density-dependent trends
of the symmetry energy can be extracted from analyzing the FOPI/GSI
data on the $\pi^{-}/\pi^{+}$ ratio in relativistic heavy-ion
collisions with various transport models~\cite{xiao09,feng10,ru11}.
In spite of this inconsistency, the theoretical uncertainty of
high-density symmetry energy is regarded to be associated with the
tensor force that originates from the exchange
terms~\cite{xu10,vi11}. In the ladder approximation, the exchange
terms can be well treated in the Brueckner theory either in the
relativistic or non-relativistic frameworks~\cite{br90,so98}. In
deed, the vacuum polarizations given by ring diagrams are absent in
the Brueckner theory. The inclusion of the ring diagrams is however
very complicated.   In this work, we do not carry on the tensor force
that appears beyond the Hartree approximation but consider the
non-nucleonic degrees of freedom  in the Hartree approximation.

The new degrees of freedom considered here are hyperons and quarks
that may appear in dense matter roughly around the density
2-4$\rho_0$, depending on the parametrization of
models~\cite{pr97,gl01,ta02,ji06,ji12}. Nuclear matter at this
density domain can be produced via heavy-ion collisions, and it
usually includes the admixture of non-nucleonic degrees of freedom.
In the past, the symmetry energy effects on these phase transitions
have been found to be very significant~\cite{ji06,di06,di11}.
However, the effects of new constituents on the symmetry energy are
seldom investigated.  It is not the aim of this work to resolve the
uncertainty of the high-density symmetry energy but to reveal the
variation of the symmetry energy in phases mixed with these new
constituents. Once the phase transition occurs, the system goes to
the mixed phase that can be theoretically constructed by virtue of
the phase equilibrium conditions, namely, the Gibbs conditions in
this work. In the mixed phase, we define the symmetry energy
according to the general expression of the energy density that is
different from that in pure nuclear matter. The paper is organized in
the following. In Sec.II, we present necessary formulas for the
nuclear symmetry energy in pure hadron and mixed phases with the
relativistic mean-field (RMF) framework. The construction of the
mixed phase of quarks and baryons are presented briefly. In Sec. III,
the numerical results and discussion are given. At last, we give a
brief summary.

\section{Formalism}
\label{RMF}

In the parabolic approximation, the energy per nucleon in isospin
asymmetric nuclear matter can be written as
\begin{equation}\label{eos1}
    {\cal{E}}/\rho=E/A=e_0(\rho)+E_{sym}(\rho)\delta^2,
\end{equation}
where $e_0(\rho)$ is the energy per nucleon in symmetric nuclear
matter, the $E_{sym}(\rho)$ is the symmetry energy, and
$\delta=(\rho_n-\rho_p)/\rho$ is the isospin asymmetry. In RMF
models, the energy density can generally be written as~\cite{ji07}
\begin{eqnarray}
  {\cal{E}}&=&\half C_\om^2\rho^2+\half C_\rho^2 \rho^2\delta^2+ \half
 \tilde{C}^2_\sg(m_N^*-M^*)^2\nnb\\
 &&+\sum_{i=p,n}
 \frac{2}{(2\pi)^3}\int_{0}^{{k_F}_i}\! d^3\!k~ E^*+{\cal{E}}_{non},
  \label{eqe1}
 \end{eqnarray}
where $C_\om=g_\om^*/m_\om^*$, $C_\rho=g_\rho^*/m_\rho^*$,
$\tilde{C}_\sg=m_\sg^*/g_\sg^*$,   $E^*=\sqrt{{\bf k}^2+{m_N^*}^2}$
with $m_N^*=M^*-g^*_\sg\sg$ the effective mass of nucleon,  $k_F$ is
the Fermi momentum, and ${\cal{E}}_{non}$ is the nonlinear meson
self-interacting energy density  specifically for the nonlinear RMF
models~\cite{nl3,tm1,ji10}. Here, the parameters with asterisks
denote the density dependence which is given in specific models. For
instance, in nonlinear RMF models, the coupling constants are
density-independent, while the meson masses are density-dependent due
to the nonlinear self-interacting terms. With Eq.(\ref{eqe1}), the
symmetry energy in the RMF models can be derived as
\begin{equation}\label{esym}
    E_{sym}=\half\frac{\pp^2({\cal E}/\rho)}{\pp\delta^2}
    =\frac{1}{2}C_\rho^2\rho +\frac{k_F^2}{6E_F},
\end{equation}
with $E_F=\sqrt{k_F^2+{m_N^*}^2}$.

For hyperon degrees of freedom, we consider only the $\Lambda$
hyperon for simplicity. In this case, Eq.(\ref{eos1}) still holds for
hyperonized matter. The nuclear symmetry energy now reads
\begin{equation}\label{esymy}
E_{sym} =\frac{1}{2}C_\rho^2\frac{\rho_N^2}{\rho_B} +
\frac{k_F^2}{6E_F}\frac{\rho_N}{\rho_B},
\end{equation}
where $k_F$ is the nucleon Fermi momentum, $\rho_N$ is the number
density of nucleons, and $\rho_B=\rho_N+\rho_\Lambda$. This formula
applies to the case of the given ratio $\rho_\Lambda/\rho_B$. The
symmetry energy is now suppressed due to the factor $\rho_N/\rho_B$.
On the other hand, as one source term of meson fields, the $\Lambda$
hyperon has led to a moderate decrease to the nucleon effective mass
and $E_F$ in the kinetic term. Together with the suppressed nucleon
Fermi momentum in $E_F$, the suppression of the kinetic term can be
partially compensated. Nevertheless, the symmetry energy eventually
turns out to be suppressed in either the baryon-density or
nucleon-density profile. In chemically equilibrated and charge
neutral matter, the particle fractions are obtained from solving
coupled equations. In this case, the nuclear symmetry energy is given
as
\begin{equation}\label{esymy1}
    E_{sym}=\frac{1}{4\delta}(\mu_n-\mu_p)\frac{\rho_N}{\rho_B},
\end{equation}
where $\mu_n$ and $\mu_p$ are the neutron and proton chemical
potentials, respectively. As the isospin asymmetry parameter $\delta$
approaches to vanishing, Eq.(\ref{esymy1}) reduces to
Eq.(\ref{esymy}).

After the hadron-quark phase transition occurs, hadrons and quarks
coexist in a mixed phase. The construction of the mixed phase is
based on the mechanical and chemical equilibriums, namely, the Gibbs
conditions which are given as~\cite{gl01}
\begin{eqnarray}\label{gibbs}
p^H&=&p^Q,\hbox{ }\mu_u=\mu_n/3-2\mu_e/3,\nnb\\
 \mu_d&=&\mu_s=\mu_n/3+\mu_e/3.
\end{eqnarray}
The pressures of nuclear and quark matter read
\begin{eqnarray}\label{presh}
 p^H&=&\half C_\om^2\rho^2_N+\half C_\rho^2 \rho^2_N\delta^2- \half
 \tilde{C}^2_\sg(m_N^*-M^*)^2 \nnb\\
 &&-\Sg_0\rho_N+
 \frac{1}{3}\sum_{i=p,n}\frac{2}{(2\pi)^3}\int_{0}^{{k_F}_i}\! d^3\!k
 ~\frac{{\bf k}^2}{E^*},\\
 p^Q&=&\sum_{i=u,d,s}\frac{2}{(2\pi)^3}\int_{0}^{{k_F}_i}\! d^3\!k
 ~\frac{{\bf k}^2}{\sqrt{{\bf k}^2+m_i^2}}-B,\nnb\\
\end{eqnarray}
where $B$ is the bag constant of the MIT model~\cite{mit} and
$\Sigma_0$ is the rearrangement term induced by the density
dependence of model parameters~\cite{ji07}. For convenient narration,
we do not include the $\Lambda$ hyperons whose addition can be
referred to Ref.~\cite{ji06}. In actual calculations, we include the
strange meson-hyperon interactions. In terms of the quark phase
proportion $Y$, the total density can be expressed as
\begin{equation}\label{density}
    \rho_B=\frac{Y}{3}\rho_Q+(1-Y)\rho_H,
\end{equation}
where $\rho_H$ is the baryon density on the hadronic level and
$\rho_Q$ is the quark density. Using Gibbs conditions, one can obtain
the quark phase proportion $Y$. The total energy density and isospin
asymmetry parameters are written as
\begin{equation}\label{eosm}
{\cal{E}}=(1-Y){\cal{E}}_H+Y{\cal{E}}_Q,\hbox{ }
\alpha=(1-Y)\delta_H+Y\delta_Q,
\end{equation}
with $\delta_H=(\rho_n-\rho_p)/\rho_N$ and
$\delta_Q=(\rho_u-\rho_d)/(\rho_u+\rho_d)$. The energy density in the
mixed phase thus depends on the $Y$. In the parabolic approximation,
the energy density can be expressed as
\begin{eqnarray}\label{eosm1}
{\cal{E}}/\rho_B&=&e_0(\rho_B,Y)+E^H_{sym}(\rho_B,Y)\delta_H^2\nnb\\
&&+E_{sym}^Q(\rho_B,Y)\delta_Q^2.
\end{eqnarray}
Because the volume fraction depends on the isospin asymmetry, we
limit the derivation of the symmetry energy $E^H_{sym}$ in symmetric
matter, namely $\alpha=0$. In this way, the nuclear symmetry energy
is defined as $E^H_{sym}=\half\pp^2({\cal E}/\rho_B)/\pp\delta_H^2$
at $\delta_H=0$, and the definition of the quark symmetry energy is
similarly given as $E^Q_{sym}=\half\pp^2({\cal
E}/\rho_B)/\pp\delta_Q^2$ at $\delta_Q=0$. Bridged by Gibbs
conditions, the quark volume fraction is model dependent and relies
on the MIT bag constant. As a result, similar dependencies on the
model and bag constant can be delivered to  the symmetry energy. As
seen from Eqs.(\ref{eosm}) and (\ref{eosm1}), the nuclear symmetry
energy may disappear as the quark fraction grows to be unity at high
densities.

\section{Numerical results and discussions}
\label{results}

\begin{figure}
\begin{center}
\vspace*{-10mm}
\includegraphics[width=8cm,height=8cm]{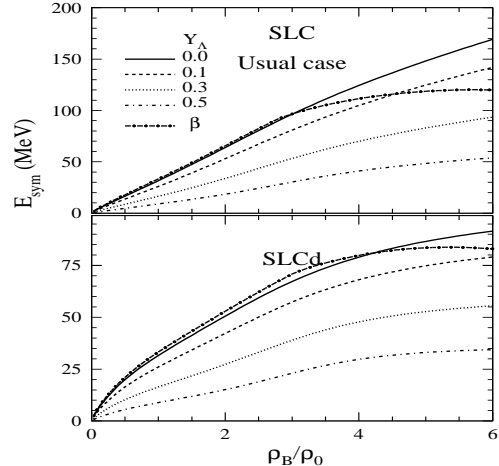}
 \end{center}
\caption{The symmetry energy as a function of density for various
$\Lambda$ fractions in the usual case (For details, see text). The
upper panel is for the results with the SLC, while the lower panel
displays the results with the SLCd.  The label $\beta$ denotes the
chemically equilibrated and charge neutral matter.} \label{fig1}
\end{figure}

\begin{figure}
\begin{center}
\vspace*{-10mm}
\includegraphics[width=8cm,height=8cm]{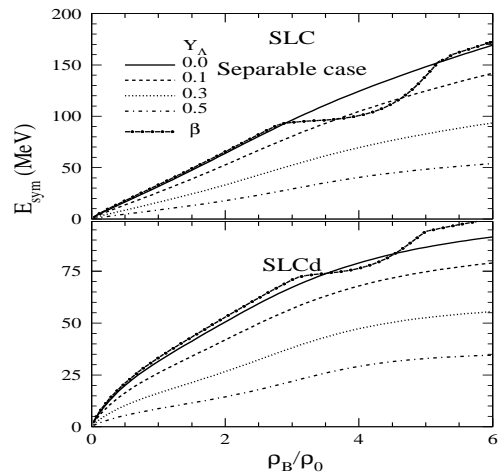}
 \end{center}
\caption{The same as shown in Fig.~\ref{fig1} but with different
density dependencies for the hyperon and nucleon potentials, denoted
as the separable case. } \label{fig2}
\end{figure}

We first display numerically the role of $\Lambda$ hyperons in
affecting the nuclear symmetry energy. In Fig.~\ref{fig1}, it shows
the density profile of the symmetry energy for various $\Lambda$
fractions with RMF models SLC and SLCd~\cite{ji07b}. The symmetry
energy for various $\Lambda$ fractions is calculated in symmetric
matter at $\delta=0$. It is shown in Fig.~\ref{fig1} that the
symmetry energy is softened clearly with the increase of the
$\Lambda$ fraction. For chemically equilibrated and charge neutral
matter, we see that the symmetry energy starts to soften once the
$\Lambda$ hyperons appear. In this case, the symmetry energy at lower
densities with the SLC is identical with that obtained with
$f_\Lambda=0$ in symmetric matter, while a small difference appears
in results with the SLCd, which is associated with the deviation in
calculating the symmetry energy at the clearly larger isospin
asymmetry in SLCd. The usual case denoted in Fig.~\ref{fig1} means
that the  hyperon-nucleon interaction has similar in-medium
properties to that of the nucleon-nucleon interaction (for details,
see Ref.~\cite{ji12}). Shown in Fig.~\ref{fig2} is the symmetry
energy for various $\Lambda$ hyperon fractions with the separable
case whose density dependence of the hyperon potential is different
from that of the nucleon potential~\cite{ji12}. Except for the
chemically equilibrated and charge neutral case, results shown in
Fig.~\ref{fig2} and \ref{fig1} are almost identical. This can be
elaborated by Eq.(\ref{esymy}) because the nuclear symmetry energy in
hyperonized matter is dominantly affected by the hyperon fraction. In
the separable case, the hyperon fraction saturates at the certain
high density, and the hyperons disappear at very high density. Thus,
in the density profile of the symmetry energy, a concave segment
forms.

\begin{figure}
\begin{center}
\vspace*{-10mm}
\includegraphics[width=8cm,height=8cm]{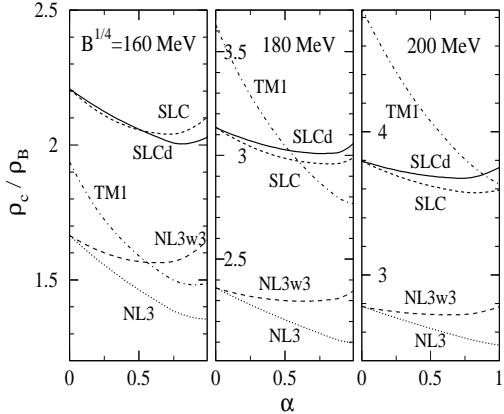}
 \end{center}
\caption{The critical density for the hadron-quark phase transition
as a function of isospin asymmetry with various RMF models and bag
constants as marked. } \label{fig3}
\end{figure}

With the increase of  density, the hadron-quark phase transition may
occur. In this work, quark matter, regarded as the free fermion gas
without interactions, is described with the MIT bag model~\cite{mit}.
For a simple example, we first deal with the phase transition without
hyperons. In a next step, we then include the hyperons. The mixed
phase consists of high-density quark matter and low-density nuclear
matter with the quark phase proportion $Y$ being obtained according
to Gibbs conditions. Here, we do not iterate the solution details
which can be found in the literature~\cite{gl01}. The critical
density of the hadron-quark transition is model-dependent and also
depends on the bag parameter clearly. In Fig.~\ref{fig3}, we depict
the critical density as a function of isospin asymmetry  for various
bag parameters and RMF models. We can observe a few characteristics
in Fig.~\ref{fig3}. First, the critical density obtained with the
TM1~\cite{tm1} is much more sensitive to the isospin asymmetry than
other models. The difference may be associated with the softening of
the vector repulsion in TM1, while other models shown in
Fig.~\ref{fig3} do not possess this softening. Second, the critical
density increases clearly with the rise of the bag constant. Third,
the softening of the symmetry energy just has a moderate effect on
the critical density. The RMF model SLCd has a softer symmetry energy
than the SLC.  The softening of the symmetry energy also occurs in
the RMF model NL3w3~\cite{ji10} as compared with the model
NL3~\cite{nl3}. One can see that the sensitivity of  the critical
density to the isospin asymmetry reduces generally due to the
softening of the symmetry energy. The exception occurs  for models
SLC and SLCd with a smaller bag parameter, see left panel in
Fig.~\ref{fig3}.

\begin{figure}
\begin{center}
\vspace*{-10mm}
\includegraphics[width=8cm,height=8cm]{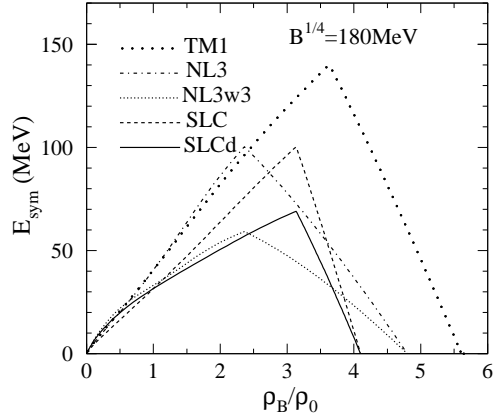}
 \end{center}
\caption{The nuclear symmetry energy as a function of density in
symmetric matter including the hadron-quark phase transition at high
densities. The reflection point from rising to dropping corresponds
to the critical density for each model. Above the critical density,
plotted is the symmetry energy $E_{sym}^H$. } \label{fig4}
\end{figure}

Besides the critical density, the quark phase proportion $Y$ also
depends on the isospin asymmetry. In this way, the symmetry energy in
the mixed phase obtained in symmetric matter can not simply be used
to predict the properties of asymmetric matter because the quark
volume fraction changes with the isospin asymmetry in asymmetric
matter. Nevertheless, the symmetry energy obtained in symmetric
matter is instructive to exhibit its variation trend in the mixed
phase. Shown in Fig.~\ref{fig4} is the nuclear symmetry energy as a
function of baryon density with the bag constant $B=(180 MeV)^4$.
Apparent decrease of the symmetry energy can be observed after the
hadron-quark phase transition occurs. With the increase of density,
the volume fraction of nucleons decreases, which causes a
straightforward reduction of the nuclear symmetry energy. As the
nucleon volume fraction reduces to zero, the nuclear symmetry energy
vanishes.

\begin{figure}
\begin{center}
\vspace*{-10mm}
\includegraphics[width=8cm,height=8cm]{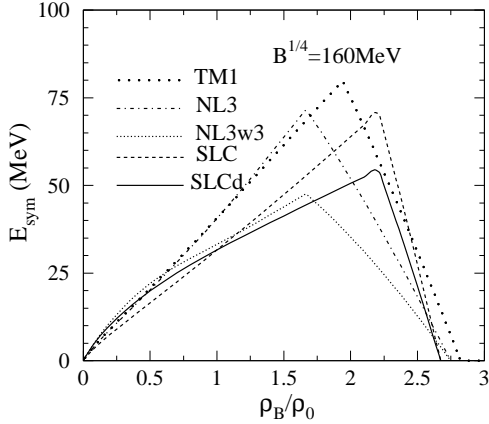}
 \end{center}
\caption{The same as shown in Fig.~\ref{fig4} but with
$B=(160MeV)^4$. } \label{fig5}
\end{figure}

For a smaller bag constant,  the critical density is also smaller,
and the decrease of the symmetry energy starts at lower densities for
various models. Shown in Fig.~\ref{fig5} is the symmetry energy with
$B=(160MeV)^4$. In this case, the critical density is around
1.6-2.2$\rho_0$ and the symmetry energy vanishes at densities below
3$\rho_0$ for all models used in the calculation. For a larger bag
constant, quite large difference in the critical density and
vanishing density of the symmetry energy exists for different models.
This can be observed by comparing Figs.~\ref{fig4} and \ref{fig5}.
For a much larger bag constant, for instance, $B=(200MeV)^4$, the
divergence becomes more appreciable. Nevertheless, once the
hadron-quark phase transition occurs, the decrease of the  symmetry
energy is definite.

\begin{figure}
\begin{center}
\vspace*{-10mm}
\includegraphics[width=8cm,height=8cm]{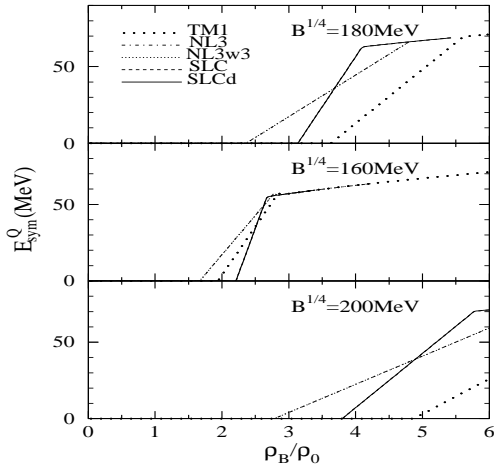}
 \end{center}
\caption{The quark symmetry energy as a function of density for
various bag constants and RMF models. The difference in the symmetry
energy exists in the mixed phase for various RMF models, and it
disappears in pure quark matter  at sufficiently high densities.  }
\label{fig6}
\end{figure}

For the quark phase, the quark symmetry energy $E_{sym}^Q$ which
reflects the cost deviating from the flavor symmetric matter starts
to have value above the critical density and increases with the rise
of the quark volume fraction in the mixed phase, as shown in Fig.~\ref{fig6}.
After the quark volume fraction quickly develops to be unity that is
a value for pure quark matter, the quark symmetry energy grows quite
slowly with the density, since without interactions only the kinetic
energy contributes to the symmetry energy.

\begin{figure}
\begin{center}
\vspace*{-10mm}
\includegraphics[width=8cm,height=8cm]{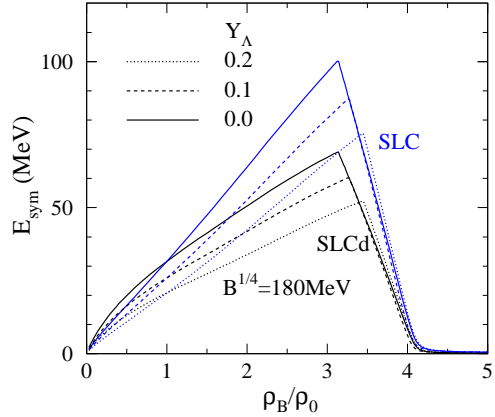}
 \end{center}
\caption{(Color online) The symmetry energy as a function of density
for different $\Lambda$-hyperon fractions in hyeronized matter  with
the hadron-quark phase transition. The RMF models SLC and SLCd are
adopted here and the bag constant is $(180MeV)^4$. } \label{fig7}
\end{figure}

With the inclusion of hyperons, the reconstruction of the chemical
equilibrium with quarks results in a larger critical density. Shown
in Fig.~\ref{fig7} is the nuclear symmetry energy in the hadronic and
mixed phases for various fractions of $\Lambda$ hyperons with RMF
models SLC and SLCd. Here, the bag constant is $B=(180MeV)^4$. The
inclusion of $\Lambda$ hyperons suppresses the symmetry energy in the
hadronic phase, consistent with those shown in Figs.~\ref{fig1} and
\ref{fig2}. In the mixed phase, the influence of $\Lambda$ hyperons
is not prominent due to the suppression factor $(1-Y)$ that decreases
quickly. For other bag constants and other RMF models, the conclusion
is qualitatively similar. Namely, the inclusion of $\Lambda$ hyperons
suppresses the symmetry energy in hadronic phase, and the decrease of
the symmetry energy starts at a larger critical density. To save
space, these numerical results are thus not displayed.

We note that the softening of the symmetry energy in the mixed phase
is mostly apparent because once the quark volume fraction can be
identified at given densities the nuclear symmetry energy would be
extracted appropriately by singling out the effect of suppression
factor $(1-Y)$. However, the determination of the $Y$ is strongly
model-dependent and  far from experimental feasibility. On the other
hand, the decreasing factor $(1-Y)$ in the mixed phase largely
suppresses the growth of the nucleonic density with the rise of the
total density. If the hyperons are taken into account, the situation
becomes more complicated. Thus, the extraction of the high-density
symmetry energy for pure nucleonic matter is not well grounded once
the hadron-quark phase transition occurs. Most likely, the
high-density symmetry energy extracted from the heavy-ion collisions
would be as soft as that presented in this work as long as a detailed
discrimination or calibration is not ready for dynamically
evolutional matter.

At last, it is worth mentioning that the appearance of new degrees of
freedom that is energetically favored usually softens the equation of
state, resulting in a significant decrease of the maximum mass of
neutron stars. Recently, the pulsar J1614-2230 was identified rather
accurately through the Shapiro delay to have a mass
$2M_\odot$~\cite{de10}, which sets up a lower limit of the maximum
mass of neutron stars.  This states that the nuclear equation of
state should not be softened significantly even with the appearance
of new degrees of freedom. A way out of this dilemma is to consider
new forms of interactions for new degrees of
freedom~\cite{ji12,be11,we12,al05,we11,bo12}. For quark matter, the
imposition of interactions can stiffen the equation of state and
hence increase the maximum mass of neutron
stars~\cite{al05,we11,bo12}. It would be interesting to investigate
whether the interactions of quarks have an effect on the nuclear
symmetry energy. This deserves subsequent work and is however beyond
the scope of the present work.

\section{Summary}
\label{summary}

We have studied the effect of $\Lambda$ hyperons and quarks on the
nuclear symmetry energy at high densities with relativistic  models.
The softening of the nuclear symmetry energy is observed either in
chemically equilibrated matter or matter with an given $\Lambda$
fraction. With the inclusion of quark degrees of freedom, we have
constructed the isospin symmetric mixed phase according to Gibbs
conditions using the RMF models and MIT bag model. The nuclear
symmetry energy obtained in the mixed phase reduces quickly with the
rise of quark volume fraction. We have recognized that the specific
softening depends on the parametrizations of models. Especially, it
has a clear dependence on the bag constant of the MIT bag model.
Nevertheless, we conclude that the effect of phase transitions is
important on the symmetry energy, and for the experimental extraction
of the symmetry energy at high densities it is significant and
necessary to take into account the effect of phase transitions.
\section*{Acknowledgement}

Authors thank Prof. X. M. Xu for useful discussions. The work was
supported in part by the National Natural Science Foundation of China
under Grant Nos. 10975033 and 11275048 and the China Jiangsu
Provincial Natural Science Foundation under Grant No.BK2009261.

\end{document}